# Specific Heat To $H_{c2}$: Evidence for Nodes or Deep Minima in the Superconducting Gap of Under- and Overdoped Ba(Fe$_{1-x}$Co$_x$)$_2$As$_2$

J. S. Kim[1], B. D. Faeth[1], Y. Wang[1], P.J. Hirschfeld[1], G. R. Stewart[1], K. Gofryk[2], F. Ronning[2], A. S. Sefat[3], K. Y. Choi[4] and K. H. Kim[4]

[1]Department of Physics, University of Florida, Gainesville, FL 32611-8440

[2]Los Alamos National Laboratory, Los Alamos, NM 87545

[3] Oak Ridge National Laboratory, Oak Ridge, TN 37831

[4] FPRD & CeNSCMR, Department of Physics and Astronomy, Seoul National University, Seoul 151-747, Republic of Korea

**Abstract:** Low temperature specific heat, C, in magnetic fields up to $H_{c2}$ is reported for underdoped Ba(Fe$_{0.955}$Co$_{0.045}$)$_2$As$_2$ ($T_c$=8 K) and for three overdoped samples Ba(Fe$_{1-x}$Co$_x$)$_2$As$_2$ (x=0.103, 0.13, and 0.15, $T_c$=17.2, 16.5, and 11.7 K respectively). Previous measurements of thermal conductivity (as a function of temperature and field) and penetration depth on comparable composition samples gave some disagreement as to whether there was fully gapped/nodal behavior in the under-/overdoped materials respectively. The present work shows that the measured behavior of the specific heat γ ($\propto$ C/T as T→0, i. e. a measure of the electronic density of states at the Fermi energy) as a function of field approximately obeys γ$\propto$H$^{0.5\pm0.1}$, similar to the Volovik effect for nodal superconductors, for both the underdoped and the most overdoped Co samples. However, for the two overdoped compositions x=0.103 and 0.13, the low field (H ≤ 10 T) data show a Volovik-like behavior of γ$\propto$H$^{0.3-0.4}$, followed by an inflection point, followed at higher fields by γ$\propto$H$^1$. We argue that within the 2-band theory of superconductivity, an inflection point may occur if the interband coupling is dominant.

I. Introduction

The structure of the superconducting gap in the iron pnictide and chalcogenide (FePn/Ch) superconductors is of central interest for understanding the underlying physics of the pairing mechanism. Of particular interest are studies of FePn/Ch superconductors where the magnetic spin density wave of the undoped compound is not yet suppressed when doping induces superconductivity, the so called 'underdoped' compounds where magnetism has been shown[1] to coexist with superconductivity. Studies of the penetration depth $\lambda$, thermal conductivity $\kappa$, NMR, tunneling and specific heat as a function of field have been used to infer the gap structure, with sometimes conflicting results.[1] One of the better characterized systems where coexistence occurs is $Ba(Fe_{1-x}Co_x)_2As_2$, where the underdoped ($0.04 \leq x \leq 0.06$, $T_c$ and $T_{SDW}$ both finite) sample has been reported as fully gapped based on thermal conductivity[2] and penetration depth[3] measurements (although ref. 4 finds nodal behavior in the c-axis thermal conductivity). In contrast, the overdoped composition (where magnetism is suppressed) has been reported as displaying nodal behavior based on the temperature (c-axis direction) and field dependence (both a- and c-axis) of the thermal conductivity data[4].

Specific heat data in field as a bulk method - less sensitive to surfaces, impurities and/or defects - of inferring the presence or absence of superconducting gap was pioneered by Moler et al.[5]. They showed that the specific heat $\gamma$ ($\propto C/T$ as $T \rightarrow 0$) in (nodal) YBCO deep in the superconducting state varied with field as $H^{0.5}$. This contrasts with $\gamma \propto H^1$ behavior found in fully gapped superconductors like $V_3Si$ or $Nb_3Sn$.[6] The $\gamma \propto H^{0.5}$ predicted by Volovik[7] comes from the effect on the electronic density of states ($\propto \gamma$) of the Doppler shift of the low-energy nodal quasiparticles in the superflow field of the vortex lattice. More recent theoretical work[8-11] has refined the early Volovik results, using $\gamma$ vs H - affected by low lying excitations in the mixed state around vortices - to make inferences about the gap structure.

Previous work[12-14] on $\gamma$(H) up to 9 T has reported that overdoped Ba(Fe$_{1-x}$Co$_x$)$_2$As$_2$ exhibits Volovik-like sub-linear behavior. Recently, the importance of measuring $\gamma$(H) over a wider field range has been made clear by results of Wang et al.[10] where they found $\gamma \propto$ H$^{0.5}$ for H<0.1 H$_{c2}$ followed by $\gamma \propto$ H$^1$ for 0.1 H$_{c2}$ < H $\leq$ 0.7 H$_{c2}$ in BaFe$_2$(As$_{0.7}$P$_{0.3}$)$_2$, which was fit to a two gap model with gap nodes or deep minima on the smaller gapped Fermi surface.

The current work presents specific heat data at low temperature as a function of fields up to close to H$_{c2}$ of the respective samples to further investigate the nodal behavior of under- and overdoped Ba(Fe$_{1-x}$Co$_x$)$_2$As$_2$. These two systems are of particular interest to compare for two reasons. As already mentioned, there are a number of (sometimes conflicting) results[1] as to whether they exhibit nodal (based on $\kappa$) or fully gapped (based on $\lambda$) behavior. Mishra et al.[15] added a new perspective to this debate showing how a system with a small gap can mimic the Volovik-like field dependence of a nodal system. Also, however, the field dependence of the definitive thermal conductivity work[4] indicates significantly different behavior between the under- and overdoped Ba(Fe$_{1-x}$Co$_x$)$_2$As$_2$. The overdoped samples (comparable to the x=0.103 composition reported here) show clear saturation of $\kappa$/T with field up to 15 T in both the a- and c-axis directions similar to that observed in the d-wave cuprate superconductor Tl-2201. In contrast, the underdoped samples show $\kappa$/T much closer to H$^1$ (gapped behavior) with, in comparison to the overdoped samples, approximately a factor of two less saturation in both directions, where saturation is define as the fraction $(\kappa_{extrap}-\kappa(15\,T))/\kappa_{extrap}$ with $\kappa_{extrap}$ equal to the $\kappa$ value extrapolated to 15 T from the low field behavior.

Thus, the present specific heat work in field provides a further investigation of the nodal properties, with a particular goal of comparing $\gamma(H)$ to the results[4] for $\kappa/T(H)$ in under- and overdoped $Ba(Fe_{1-x}Co_x)_2As_2$.

## II. Experimental

Single crystals of $Ba(Fe_{1-x}Co_x)_2As_2$, x=0.045 and 0.103, were grown[16] out of FeAs flux with a typical size of about $2\times1.5\times0.2$ mm$^3$. They crystallize in well-formed plates with the [001] direction perpendicular to the plane of the crystals. In order to sharpen the superconducting transition after synthesis the samples were annealed[17] for two weeks at 800 °C in vacuum, with resulting superconducting transition temperatures, $T_c$ of 8.0 K and 17.2 K for x=0.045 and 0.103 respectively. After the initial characterization of C(H) in the present work of these samples, single crystals of a further two compositions, x=0.13 and 0.15, were grown and annealed in a similar fashion, resulting in $T_c$ values of 16.5 and 11.7 K respectively. One important parameter in the FePn/Ch superconductors that pertains to quality of sample and amount of disorder is the specific heat $\gamma$ (C/T for T→0) in zero field. For example, in the P-doped $BaFe_2As_2$ sample of ref. 10, this so-called 'residual' $\gamma$ is 1.7 mJ/molK$^2$, which as discussed in ref. 1 is typical of high quality samples to date. The residual $\gamma$ values for the four Co-doped samples measured in the current work are approximately 2 (x=0.045), 4 (x=0.103), 9 (x=0.13) and 6.5 (x=0.15) mJ/molK$^2$. Such FePn/Ch samples do exhibit some sample dependence in their residual $\gamma$ values. In a work on samples from the same batches as the two lower Co-concentration samples measured for the current work, Gofryk et al.[17] found values of 1 and 4 mJ/molK$^2$ for the same two compositions respectively.

Specific heat was measured up to 14 T in house, and up to 26 T at the NHMFL at Tallahassee, using our established time constant method[18].

III. Results and Discussion

The as-measured specific heat divided by temperature (C/T) vs temperature as a function of field for the overdoped Ba(Fe$_{0.897}$Co$_{0.103}$)$_2$As$_2$ sample is shown in Fig. 1a. The C/T data taken in the superconducting magnet up to 14 T and down to 0.4 K show an upturn at lower temperatures due to the nuclear hyperfine contribution (splitting of the nuclear moment energy levels) in an applied magnetic field. This nuclear contribution to the specific heat in field is primarily from the $^{75}$As, but - despite the relatively low concentrations of Co used in this sample - comes also about 40% from the Co at the overdoped concentrations. Thus, the C/T data at 14 T and 0.43 K for Ba(Fe$_{0.897}$Co$_{0.103}$)$_2$As$_2$ have an ≈ 2.8 mJ/molK$^2$ nuclear contribution. Once this contribution is subtracted (shown in Fig. 1b), the slight (sample dependent) anomaly in C/T visible in the zero field data around 0.7 K in Ba(Fe$_{0.897}$Co$_{0.103}$)$_2$As$_2$ (which is absent in the Ba(Fe$_{0.955}$Co$_{0.045}$)$_2$As$_2$ and Ba(Fe$_{0.97}$Co$_{0.13}$)$_2$As$_2$ samples and less marked in the Ba(Fe$_{0.85}$Co$_{0.15}$)$_2$As$_2$ sample discussed below) remains. Such sample dependent anomalies have been seen in other iron pnictide samples.[19]

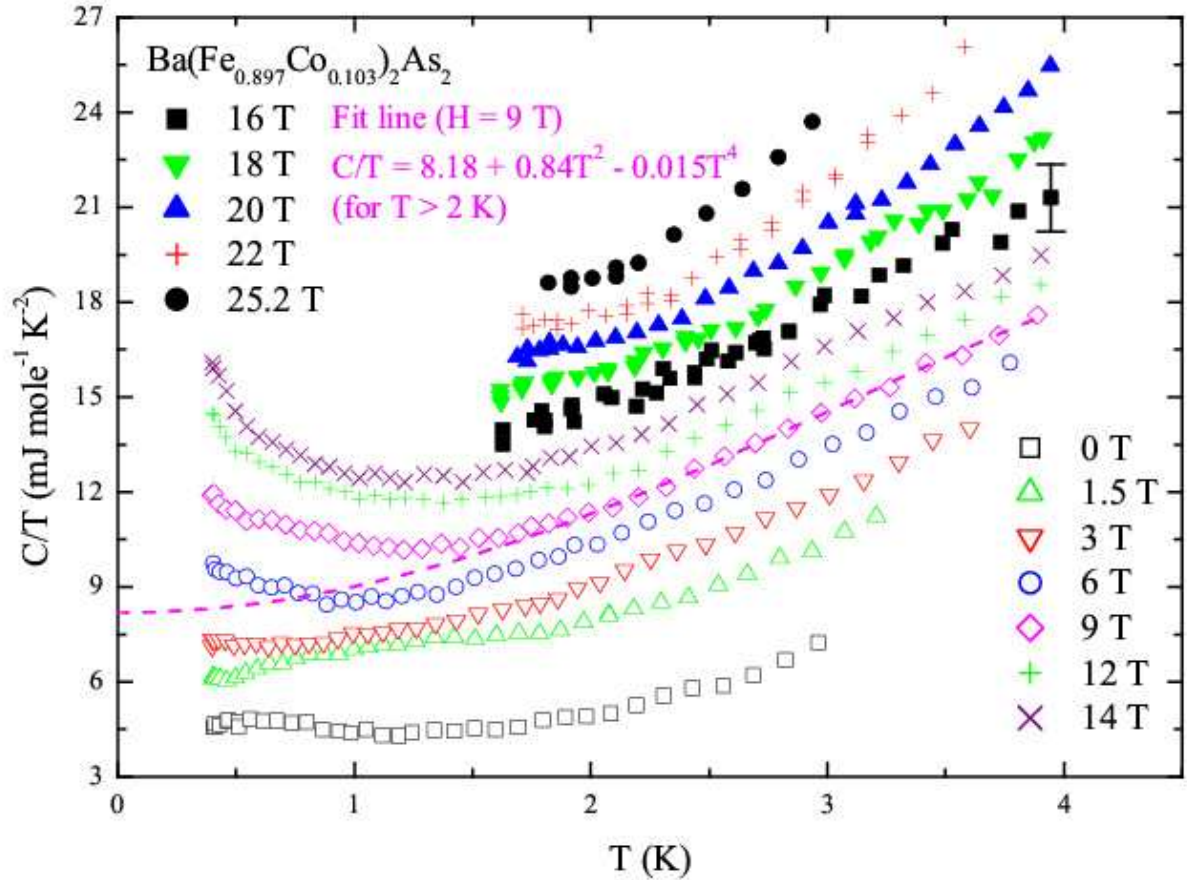

Fig. 1a: (Color online) Specific heat divided by temperature, C/T, vs temperature for annealed single crystals of Ba(Fe$_{0.897}$Co$_{0.103}$)$_2$As$_2$ as a function of field, with field aligned perpendicular to the a-b plane. The data for H≤14 T were taken down to 0.4 K in a superconducting magnet. These data have less scatter than the data taken for H≥16 T and T≥1.7 K in the normal magnet at NHMFL in Tallahassee. The absolute error bar is ≲ ±5 % for all the data in this work, approximately the size of a data point for lower fields and temperatures, while the precision of the data is approximately ±2 %. Note the upturn in C/T at low temperatures due to the field splitting of the nuclear levels. The dashed line is the three term polynomial fit to the 9 T data shown as a representative example as discussed in the text. All the polynomial fits for the data shown in this figure were for T>2 K to avoid both the low temperature anomaly and the upturn due to the nuclear hyperfine field splitting contribution. The apparent increase of the slope of C/T vs T around 3 K for H>14 T compared to the lower field data is possibly caused by the suppressed superconducting transition coming down in temperature with increasing field, since H$_{c2}$ is approximately 27 T.

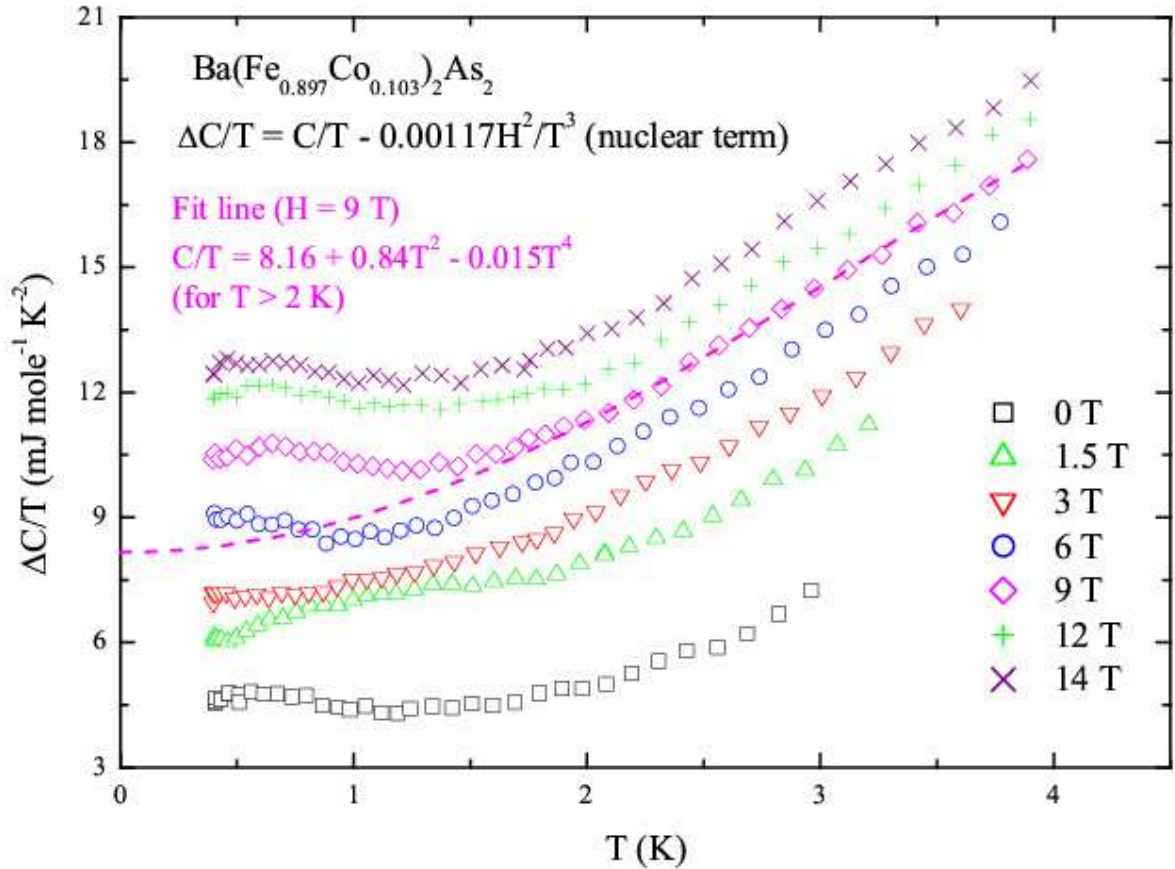

**Fig. 1b:** (Color online) Specific heat corrected for the nuclear hyperfine contribution in field divided by temperature, C/T, vs temperature for annealed single crystals of Ba(Fe$_{0.897}$Co$_{0.103}$)$_2$As$_2$ as a function of field, with field aligned perpendicular to the a-b plane. The polynomial fit line is the same as discussed for Fig. 1a.

The as-measured specific heat divided by temperature (C/T) vs temperature (note that the lowest temperature of measurement is 1.6 K) as a function of field for the overdoped Ba(Fe$_{0.87}$Co$_{0.13}$)$_2$As$_2$ sample is shown in Fig. 1c.

In the present work, in order to make clear that the variation of C/T at low temperatures that we observe is not affected by the presence or absence of these sample dependent low temperature anomalies[19], we present and compare two methods of analyzing the low temperature specific heat data. The straightforward determination is simply to fit the C/T data above the anomalies to a three term polynomial, $C/T = \gamma + \beta T^2 + \delta T^4$, (as shown by a dashed line in each figure for one field as an example) over a temperature range of several Kelvin, and utilize this fit to determine γ (C/T as T→0 is defined as γ). Such an extrapolated determination of γ as a function of H from the data in

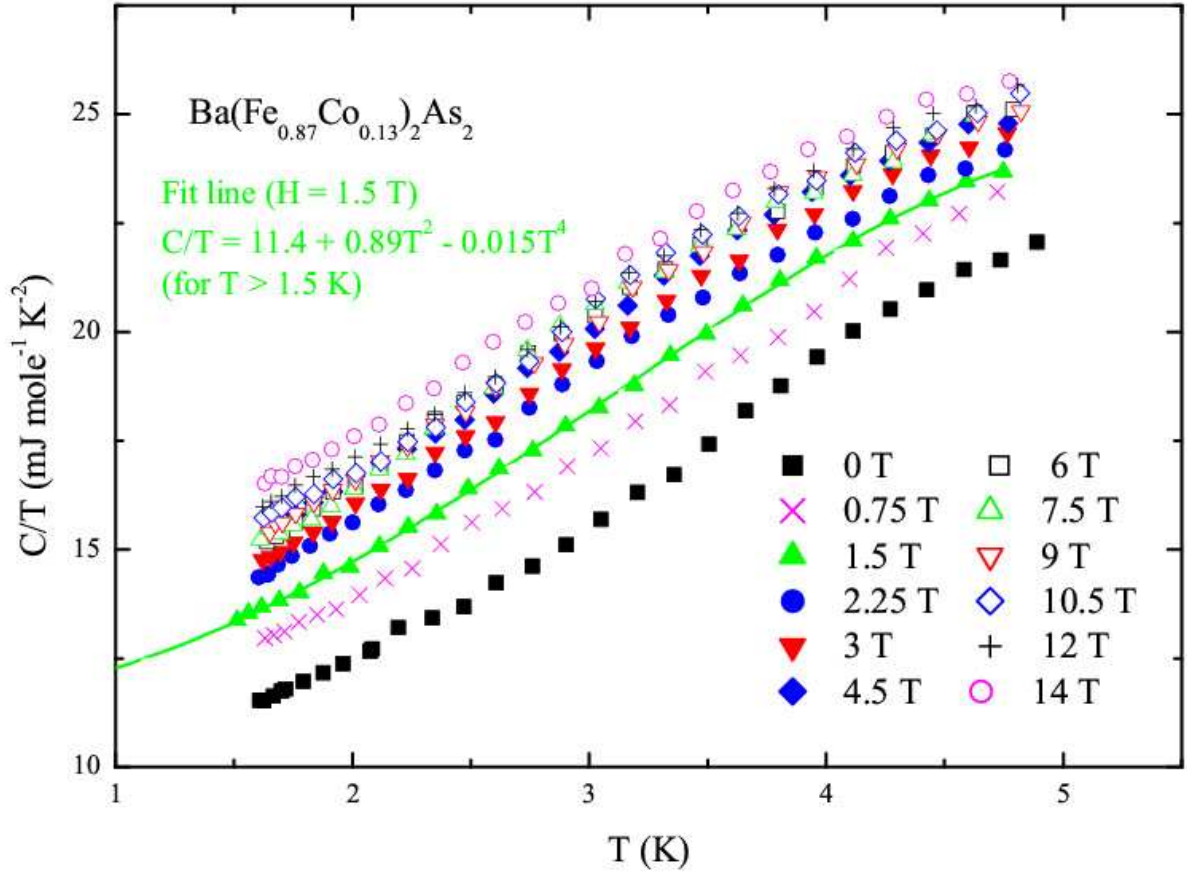

Fig. 1c: (Color online) Specific heat divided by temperature, C/T, vs temperature for annealed single crystals of Ba(Fe$_{0.87}$Co$_{0.13}$)$_2$As$_2$ as a function of field, with field aligned perpendicular to the a-b plane. These data for H≤14 T were taken down to 1.6 K in a superconducting magnet. The data taken (6-8 points per field) for H≥14 T in the normal magnet at NHMFL in Tallahassee (not shown), were only taken in the vicinity of 2 ± 0.1 K in order to maximize the data taken in the limited measurement time, providing C/T (2 K) vs H as an indication of the variation of N(0) with higher field as discussed in the text. The dashed line is the three term polynomial fit to the 1.5 T data shown as a representative example as discussed in the text. All the polynomial fits for the data shown in this figure were for T>1.5 K, i. e. down to the lowest temperature of measurement. Note the suppressed zero in the figure for the temperature axis. H$_{c2}$ for this composition is approximately 26-27 T, so that the sample at 14 T is still well in the superconducting state.

Figs. 1a and 1b for Ba(Fe$_{0.897}$Co$_{0.103}$)$_2$As$_2$ and in Fig. 1c for Ba(Fe$_{0.87}$Co$_{0.13}$)$_2$As$_2$ is presented in Fig. 2.

A second method to track the behavior of C/T (T→0) is to determine the behavior of C/T at 2 K (obtained by fitting the data around 2 K to smooth out any scatter) vs H as being representative of the behavior of γ. This second method, which is less data intensive (thereby requiring less measurement time in the delimited time environment of the NHMFL), is as shown in Fig. 2 also a fair representation of γ (≡C/T as T→0) vs H.

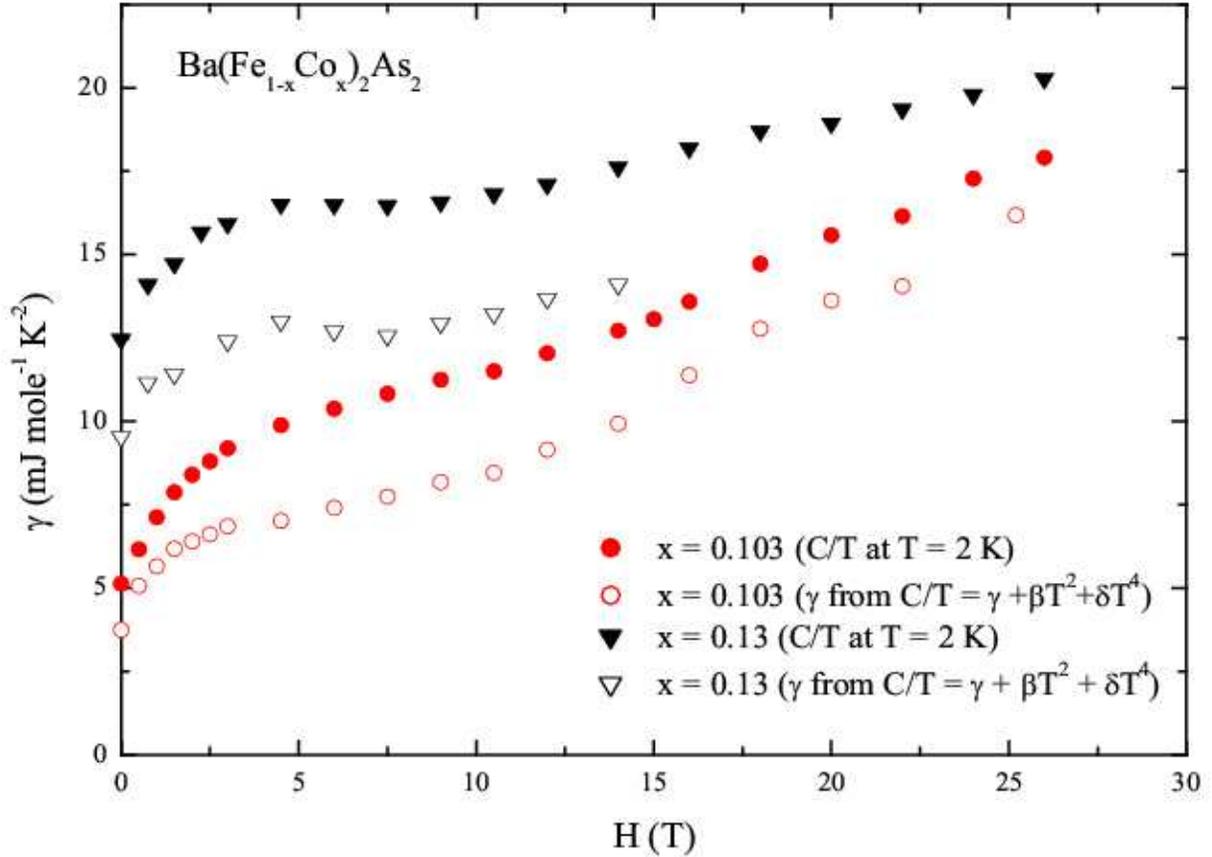

Fig. 2: (Color online) γ (open symbols) is plotted vs H for annealed single crystals of Ba(Fe$_{0.897}$Co$_{0.103}$)$_2$As$_2$ and Ba(Fe$_{0.87}$Co$_{0.13}$)$_2$As$_2$ determined from a polynomial fit to the respective data above 2 K. Secondly, C/T at 2 K (solid symbols), determined by a fit of data around 2 K to smooth the results, for both compositions is plotted vs field as a second

indication of $\gamma$ vs H.  (Sufficient data over a wide enough temperature range for x=0.13 for H>14 T to fit $\gamma$ were not measured.)  Both methods for determining the behavior of the electronic density of states at the Fermi energy, N(0), for x=0.103 track each other's variation with H up almost to $H_{c2}$ (27 T).  Note that for both x=0.103 and 0.13 the Volovik-like behavior indicative of nodes or deep minima in the gap function at low field, where $\gamma \propto H^{0.4}$, crosses over to $\gamma \propto H^1$ above around 10 T.

This procedure also avoids the temperature region where there is an anomaly in C/T, provides a good measure of $\gamma$ vs H (since C/T as T→0 and C/T (2 K) are closely related) and has relatively low scatter.  Since for measurements made in the high field, normal state magnet at NHMFL in Tallahassee above 14 T the temperature range extended only down to ≈ 1.7 K, this method of fitting the data around 2 K to arrive at a smoothed value for C/T (2 K) as representative of $\gamma$ provides a consistent measure over the entire field range as shown in Fig. 2.  Before these results for $\gamma$ vs H for Ba(Fe$_{1-x}$Co$_x$)$_2$As$_2$ , x=0.103 and 0.13 are discussed, we present the data for Ba(Fe$_{1-x}$Co$_x$)$_2$As$_2$ , x=0.045 and 0.15, first so that all the $\gamma$ vs H results can be discussed together.

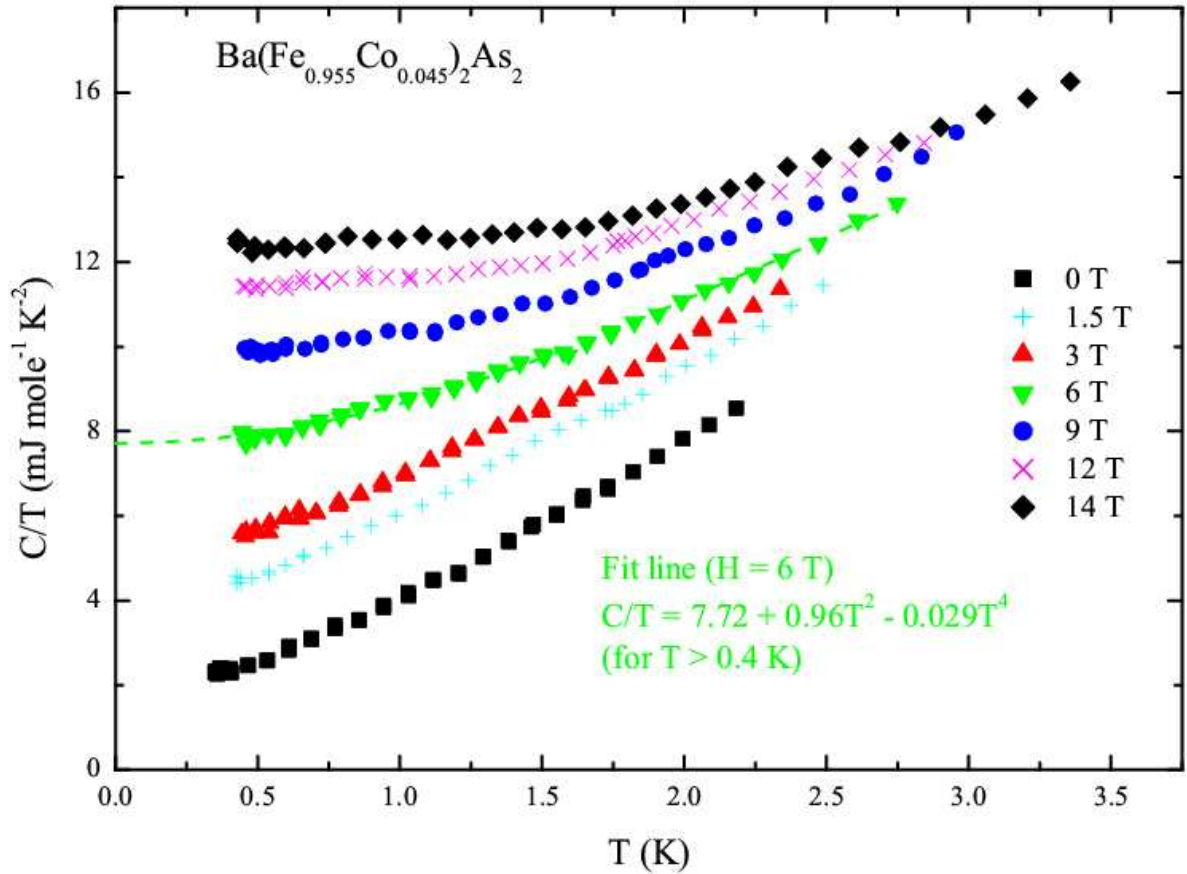

**Fig. 3a:** (Color online) Specific heat divided by temperature, C/T, vs temperature for annealed single crystals of Ba(Fe$_{0.955}$Co$_{0.045}$)$_2$As$_2$ as a function of field. Data taken at NHMFL down to 1.7 K (not shown) for H=16, 18 and 20 T show that $\gamma$ is saturated at those higher fields, so that H$_{c2}$ for this composition is approximately 15-16 T. The dashed line is the three term polynomial fit to the 6 T data shown as a representative example as discussed in the text. All the polynomial fits for the data shown in this figure were for T>0.4 K. Since H$_{c2}$ for this composition is only approximately 15 T, increasing the field to 14 T causes a noticeable 'flattening' of C/T vs T compared to the higher H$_{c2}$ samples in Figs. 1a and 1c.

The C/T data in fields to 14 T and down to 0.4 K for Ba(Fe$_{0.955}$Co$_{0.045}$)$_2$As$_2$ are shown in Fig. 3a, with data up to 14 T and down to 0.4 K for Ba(Fe$_{0.85}$Co$_{0.15}$)$_2$As$_2$ shown in Fig. 3b. As can be seen by comparing to the data in Fig. 1a for Ba(Fe$_{0.897}$Co$_{0.103}$)$_2$As$_2$, there is no (Co$_{0.045}$) or only a slight (Co$_{0.15}$) low temperature (sample dependent[19]) anomaly to hamper extrapolation of the low temperature data to T→0. Also, the nuclear hyperfine field caused upturn in C/T evident already in 6 T at 0.4 K for the $Co_{0.103}$ and $Co_{0.15}$ compositions is absent (except perhaps for the lowest temperature in 14 T) in the $Co_{0.045}$ composition, despite the fact that, based on the percentages of the contributing nuclei in the sample, the hyperfine upturn in the smaller Co-concentration sample would only be reduced by ≈25 %.

The reason for this apparent contradictory difference in observed nuclear hyperfine upturns in C/T is that the nuclear spins' coupling to the lattice is dependent on the electron density at the Fermi energy. It is well known[20] that the nuclear spin-lattice relaxation (or coupling) is *reduced* in the superconducting state (actually to zero for, e. g., the nuclear quadrupole levels split by electric field gradients in In) by the reduction of the electronic density of states at the Fermi energy. Further, in superconducting underdoped $Ba(Fe_{0.955}Co_{0.045})_2As_2$ this electron density is reduced even more by the spin density wave transition, which removes additional electronic density of states at the Fermi energy. Thus, the reduced upturn in C/T at low temperatures from nuclear levels split by the applied magnetic field in the underdoped sample vis-à-vis the overdoped samples is consistent with the expected relative reductions in the electronic density of states.

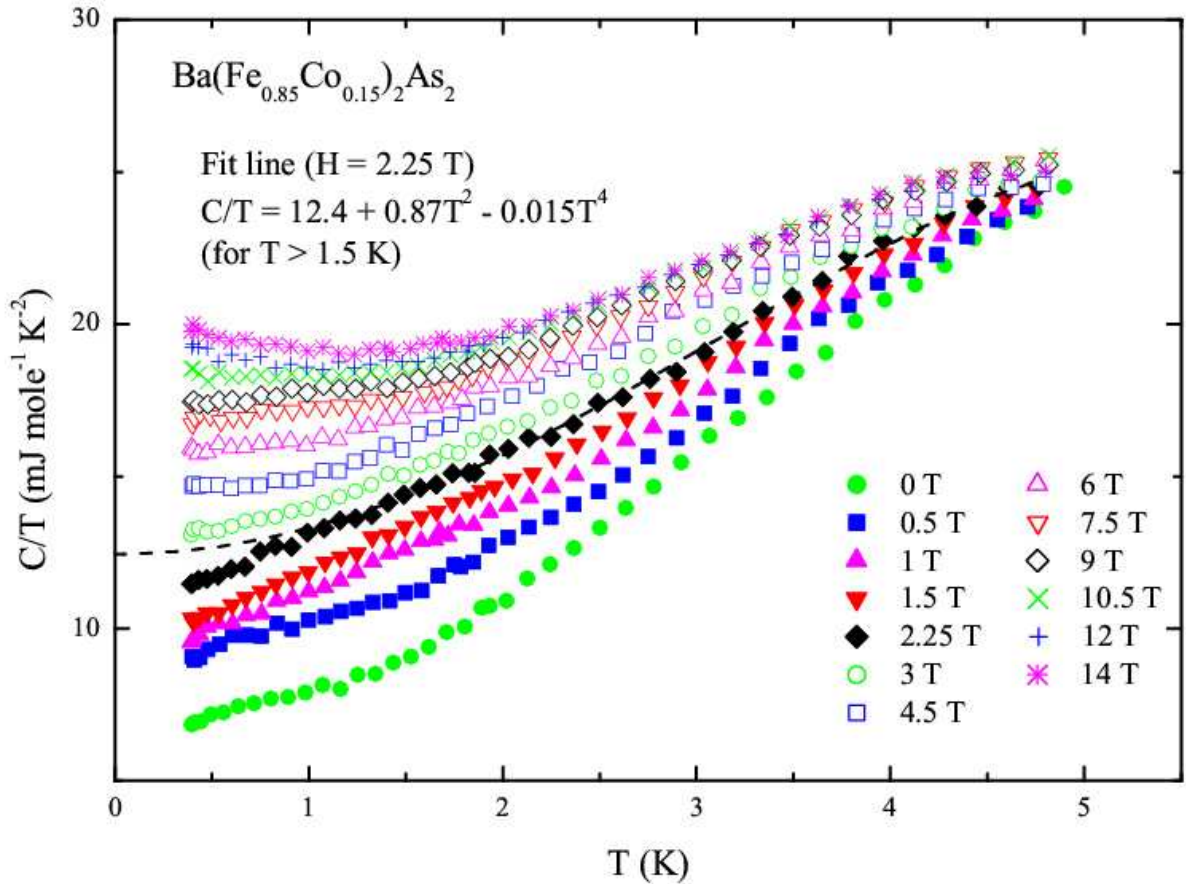

Fig. 3b: (Color online) Specific heat divided by temperature, C/T, vs temperature for annealed single crystals of Ba(Fe$_{0.85}$Co$_{0.15}$)$_2$As$_2$ as a function of field measured in a superconducting magnet. (No data at higher field were taken.) The dashed line is the three term polynomial fit to the 2.25 T data shown as a representative example as discussed in the text. All the polynomial fits for the data shown in this figure were for T>1.5 K to avoid both the low temperature anomaly and the upturn due to the nuclear hyperfine field splitting contribution. The apparent 'flattening' of the C/T vs T with increasing field compared with, e. g., the data in Figs. 1a and 1c is because x=0.15 has an H$_{c2}$ not more than 5-6 T higher than the 14 T maximum field shown here.

Fig. 4 shows γ vs H obtained from a three term fit to the C/T data in Figs. 3a and 3b for Ba(Fe$_{0.955}$Co$_{0.045}$)$_2$As$_2$ and Ba(Fe$_{0.85}$Co$_{0.15}$)$_2$As$_2$, as well at C/T (2 K) vs H. Literature data[14] up to 9 T on a Ba(Fe$_{1-x}$Co$_x$)$_2$As$_2$ sample comparable to the present work's x=0.15 composition[21] are shown for comparison.

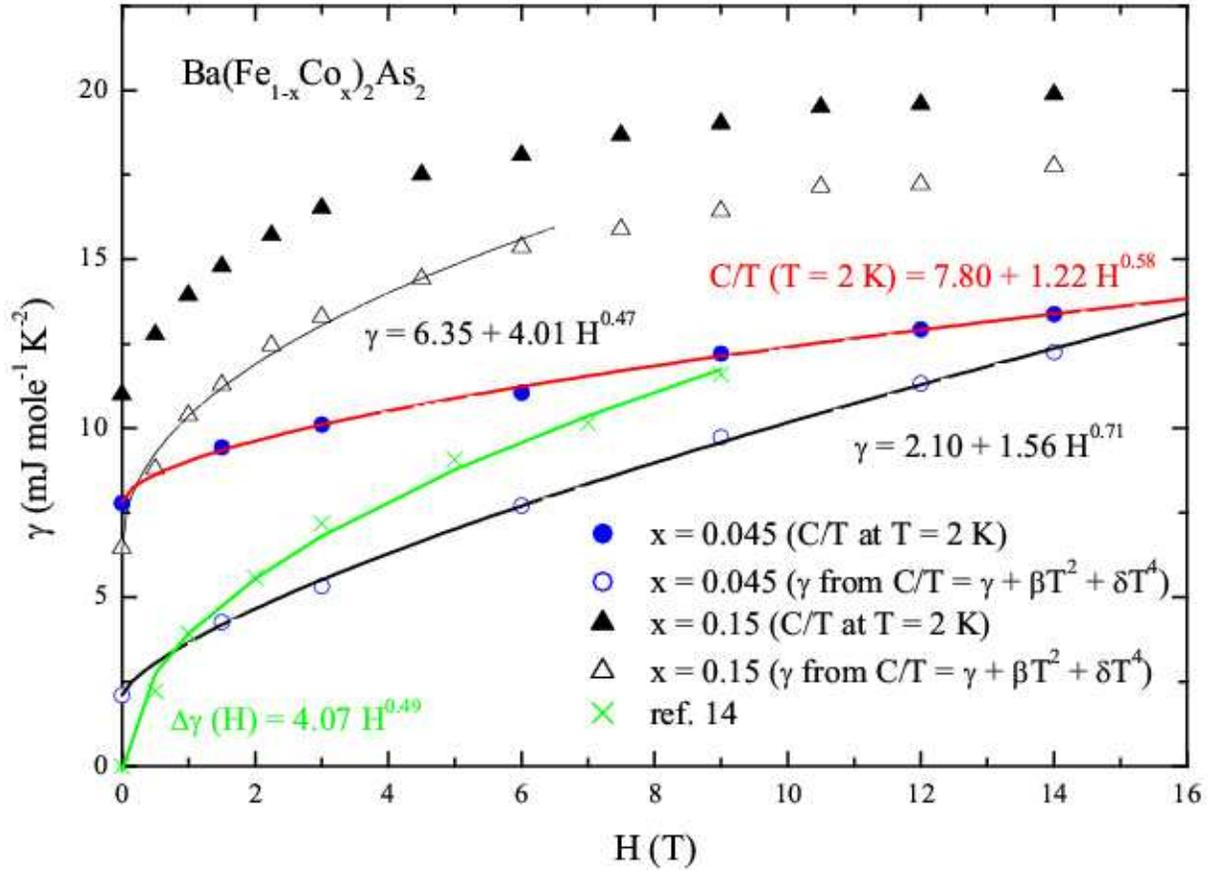

**Fig. 4:** (Color online) C/T at 2 K (determined by a fit of data around 2 K to smooth the results) vs field as an indication of $\gamma$ vs H for annealed single crystals of Ba(Fe$_{0.955}$Co$_{0.045}$)$_2$As$_2$ and Ba(Fe$_{0.85}$Co$_{0.15}$)$_2$As$_2$ as a function of field as well as $\gamma$ determined by a three term fit to the low temperature data from Figs. 3a and 3b (which have either no, or only a slight, low temperature anomaly respectively to hinder the determination of $\gamma$). Note that both methods for determining $\gamma$ vs H are in good qualitative agreement with each other. Note the change in the rate of rise in the H<6 T data for the x=0.15 data (accentuated by the fit line) vs the higher field data. Published $\gamma$ data up to 9 T from ref. 14 (see also ref. 12) (H || c-axis) using annealed (800 °C for 20 days) crystals of a composition[21] which, based on its T$_c$ (12.5 K), is similar to the x=0.15 sample of the present work (T$_c$=11.7 K) are included for comparison. These published data are shifted by their zero field residual gamma values, $\Delta\gamma(H)=\gamma(H) - \gamma(H=0)$, to make the data easier to view. Note that the magnitude of $\Delta\gamma$ with 9 T applied field, and also the deviation from a single power law fit around 3-7 T, of these ref. 14 data are similar to the x=0.15 data of the present work.

**Discussion of γ vs H; Comparison with Theory; Implications for the Nodal Structure:**

We have presented low temperature determinations of the variation of γ with field in four samples of Ba(Fe$_{1-x}$Co$_x$)$_2$As$_2$, with x=0.045, 0.103, 0.13, and 0.15. As an additional safeguard against the possible influence of low temperature (≈ 0.7 K) anomalies in the specific heat visible in the x=0.103 and 0.15 samples, C/T (2 K) was also presented for comparison. Except for a somewhat different field dependence for γ vs H for x=0.045 (γ∝H$^{0.7}$ and C/T (2 K)∝H$^{0.6}$), the two different means of tracking the change in the density of states at the Fermi energy, N(0), with field agree well.

What these data reveal is that for x=0.103 and 0.13, there appears to be a Volovik effect (i. e. γ varies ≈ H$^{0.4}$) at low fields, indicative of gap nodes or a deep gap minimum on at least one of the bands, followed surprisingly by an *inflection point* around 10 T, followed by linear in H behavior up to the highest field of measurement. In general, one expects that the various bands with superconducting gaps in a material should couple rather strongly[22], and that the separate behaviors in the individual bands should merge together to present an average curvature of γ with field over the whole field range. This is what is indeed observed in the present work for x=0.045. For x=0.15, there is a change visible in the curvature of γ with H between the low field (H<6 T) and higher field data as shown in Fig. 4, but without the distinct inflection point seen for x=0.103 and 0.13 in Fig. 2.

To understand the possible origins of an inflection point in the magnetic field dependent density of states, we consider the theory of a 2-band superconductor[23-26] as a crude approximation to the actual multiband Fe-based superconductor consisting of several hole and electron pockets. The first obvious possibility is that in a situation where the two gaps are weakly coupled, one will be suppressed nearly to zero by the application of a field at a value close to the smaller critical field for hypothetical zero interband coupling strength; in this case one would expect a rather rapid crossover to a saturated behavior with field for N(0;H) in the band hosting the small gap. We have investigated this possibility extensively using the one-vortex, 2-band framework of Ref. 10 suitable for the low-field regime, and have found no satisfactory solution with an inflection point within the range of applicability of the theory. A more appealing approach, particularly given the other evidence for strong interband coupling[22], is to note that an inflection point is possible if one of the gap magnitudes is quite large relative to the isotropic BCS value of 1.76 Tc. Since the inflection points occur at somewhat higher fields (of order 30% of Hc2), we adopt the simple Pesch approximation[27] for the zero energy DOS N(0;H) spatially averaged over the Abrikosov vortex lattice,

$$N^{\alpha}(0;H) = N_0^{\alpha} \int \frac{d\theta}{2\pi} \frac{1}{\sqrt{1+f_{\alpha}(\theta,H)}}; \quad f_{\alpha}(\theta,H) = \frac{4\Phi_0 \Delta_{\alpha}^2(\theta,H)}{\pi \hbar^2 H v_{F,\alpha}^2},$$

where $\alpha=1,2$ indexes the two bands, $N_0^{\alpha}$ is the normal state density of states, $\Delta_{\alpha}$ is the gap, and $v_F^{\alpha}$ is the Fermi velocity on band $\alpha$. Here $H$ is the applied field and $\Phi_0$ is the flux quantum. For isotropic s-wave gaps, it is easy to show that the normalized density of states $N(0;H)/N_0^{\alpha}$ has an inflection point when $f_{\alpha}'^2 = 2 f_{\alpha}''(1+f_{\alpha})/3$ (prime means derivative with respect to field H), and we expect that similar conditions hold for anisotropic gaps. Evaluation of the isotropic case gives an inflection point if

$$\eta \equiv \frac{4\Phi_0 \Delta_{\alpha}^2(H=0)}{\pi \hbar^2 H_{c2} v_{F,\alpha}^2} > \frac{4}{3},$$

so it is seen that a large gap makes such a point more likely. The inflection point occurs at

$$\frac{H_{\text{infl}}}{H_{c2}} = \frac{\eta}{4(\eta-1)}; \quad \eta = \frac{2}{e^{\gamma}}(c\zeta)^{\frac{2}{c^2\zeta+1}},$$

where $\eta$ is also obtained from the Pesch approximation in the vortex lattice state, $c=\Delta_e/\Delta_h$, $\zeta=N_0^e/N_0^h$, and $\gamma=$Euler's constant. Within 2-band BCS theory, it is difficult to get large values of $\Delta/T_c$, but within Eliashberg theory it is even possible to have two such large values.[25] To show a concrete case which exhibits the inflection, we plot in Fig. 5 a calculation within the Pesch approximation for the vortex lattice, where the gap magnitudes have been solved self-consistently within the 2-band model. Note that within the Pesch approximation, some of the downward curvature with field at low field is artificial and is obtained also for a fully gapped s-wave state.[28] A more detailed, fully self-consistent numerical analysis could capture both the correct Volovik physics as well as the inflection and crossover to $H_{c2}$. In any case the current approximation scheme appears to work very well for the anisotropic superconducting gaps over a wide range of field strength.

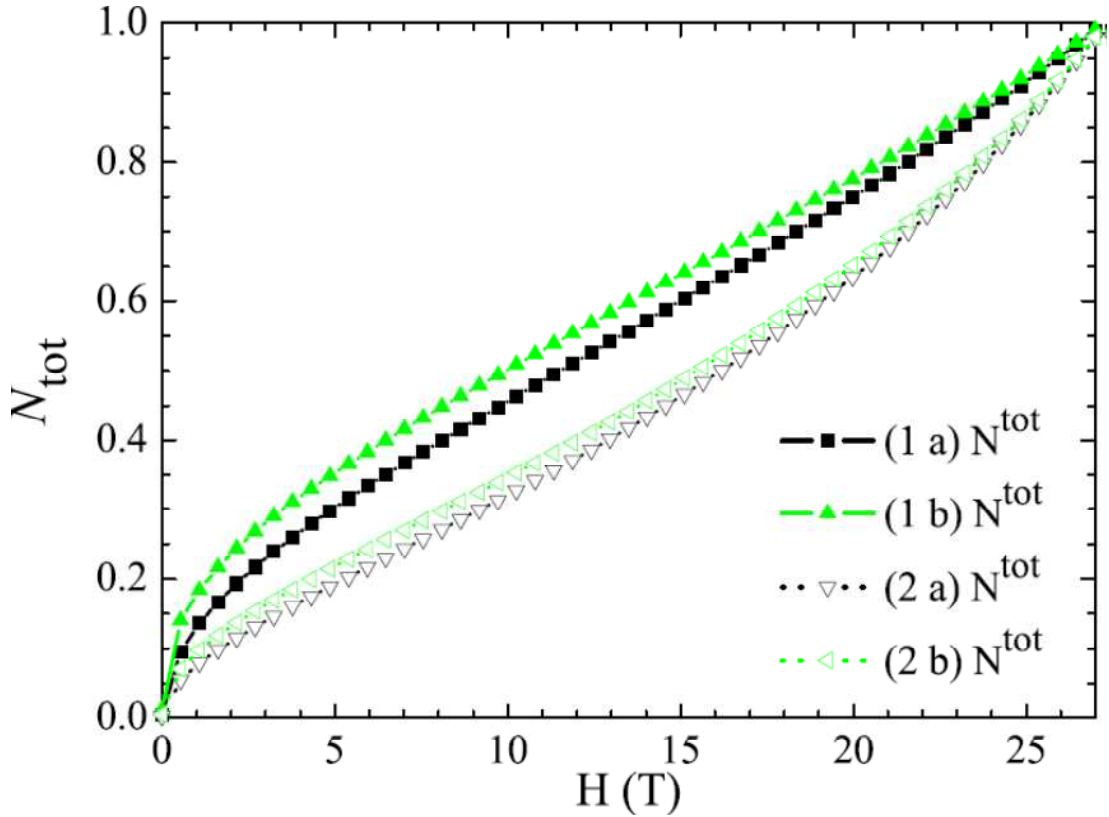

**FIG. 5:** (Color online) The normalized total zero energy density of states in the vortex lattice states by the theoretical calculation[27] with self-consistently determined gaps $\Delta_0^{e,h}(H)$ (subscript '0' means zero temperature) along the electron and hole pockets. Results (1a,b) and (2a,b) correspond to two different sets of coupling matrices used in the calculation, both with strong inter-band coupling strength, while the normal density of states ratio used in (1a,b) (solid symbols) is $N_0^h/N_0^e = 6.7$ and in (2a,b) $N_0^h/N_0^e = 33$. (a) (black symbols) and (b) (green symbols) correspond to two gap-models: (a) both $\Delta_0^{e,h}(H)$ are isotropic; (b) $\Delta_0^h(H)$ is isotropic and $\Delta_0^e(H)$ has the momentum dependence assumed to be $(1+1.3\cos 2\theta)$.

Thus, the present work presents two conclusions. The overdoped Ba(Fe$_{1-x}$Co$_x$)$_2$As$_2$, as expected[1] from other measurements (e. g. thermal conductivity) shows evidence for Volovik-like behavior in the non-linear low field variation of γ with H. As argued earlier, this need not indicate gap nodes, but is also consistent with very small gap minima over the field range observed. In addition, for the first time, the present work gives specific heat evidence that at least one and possibly two of bands exhibit gaps with $\Delta/T_c$ significantly larger than the BCS value. Because of the rather weak inflection found in the theory relative to the data in at least some of the samples, together with the rather extreme values of the density of states ratio $\zeta=N_0^e/N_0^h$ required to produce an inflection point, we regard the theoretical explanation as suggestive but not conclusive. The exact origin of the inflection point must still be regarded as an open question.

It is interesting to compare the present work with the previously published results in $BaFe_2(As_{0.7}P_{0.3})_2$, where γ varies as $H^{0.5-0.6}$ up to around 4 T, and then varies linearly ($H^1$) up to 35 T (=2/3 $H_{c2}$). In that work[10], it is stated that "the unusually small range of Volovik-type behavior, followed by a large range of linear-H behavior, is due to the small gap and weak nodes on the small mass (presumably electron) sheet," with the larger mass hole sheets believed to be fully gapped. At least for the data in $BaFe_2(As_{0.7}P_{0.3})_2$ that were published[10], crossover between the low and high field behaviors seemed more gradual and the crossover took place at a much smaller fraction of $H_{c2}$. While the stronger inflection point in the Co-doped system may imply that the interband pairing in this system is stronger than in the P-doped system, such a conclusion would clearly require further analysis.

In conclusion, the specific heat of both under- and overdoped $Ba(Fe_{1-x}Co_x)_2As_2$, at low fields exhibits a Volovik-like non-linear behavior of γ with H which is consistent with the temperature dependence of thermal conductivity[4] and indicative of nodes or deep minima in at least one band. An inflection point in $C(H)/T$ for two samples near optimal doping is argued to be evidence for strong interband coupling, and that at least one of the gaps is larger than allowed by BCS theory.

**Acknowledgements:** Work at Florida performed under the auspices of the United States Dept. of Energy, Office of Basic Energy Sciences, contract no. DE-FG02-86ER45268 and DE-FG02-05ER46236. Work at Los Alamos National Laboratory was performed under the auspices of the U. S. Dept. of Energy, Office of Basic Energy Sciences, and supported in part by the Los Alamos Laboratory Directed Research and Development program. Work at Oak Ridge National Laboratory was performed under the auspices of the U. S. Dept. of Energy, Office of Basic Energy Sciences, Materials Sciences and Engineering Division. Work at Seoul National University by K.-Y. Choi was supported by Basic Science Research Program through the National Research Foundation of Korea (NRF) funded by the Ministry of Education, Science and Technology(Grant No. 2012008233). Work at Seoul National University by K. H. Kim was supported by the National Creative Research Initiative (2010-0018300) and the Fundamental R&D Program for Core Technology of Materials of MOKE. Data above 14 T were measured at the NHMFL in Tallahassee, which is supported by the NSF.